\documentclass{article}

\usepackage[utf8]{inputenc}
\usepackage{epsfig}
\usepackage{times}
\usepackage{color}
\usepackage{amsfonts}
\usepackage{amssymb}
\usepackage[cmex10]{amsmath}
\usepackage{color}
\usepackage{xcolor}

\title{The effect of heterogeneous distributions of social norms on the spread of infectious diseases}
\author{Daniele Vilone$^{1,2}$, Giulia Andrighetto$^{1,3,4}$\\
\ \\
$^1$LABSS-ISTC, National Research Council, Rome, Italy; \\
$^2$GISC-Dep.to de Matem\'aticas, Universidad Carlos III de Madrid, Spain \\
$^3$M\"alardalen University, Vasteras, Sweden\ \\
$^4$Institute for Future Studies, Stockholm, Sweden
}

\date{March 29th, 2022}

\begin{document}

\maketitle

\ 

\begin{abstract}

\noindent The emergence due to the outbreak of the COVID-19 disease, caused by the SARS-CoV-2 virus, suddenly erupted at the beginning of 2020 in China and soon spread worldwide. This has caused an outstanding increase on research about the virus itself and, more in general, epidemics in many scientific fields. In this work we focus on the dynamics of the epidemic spreading and how it can be affected by the individual variability in compliance with social norms, {\it i.e.}, in the adoption of health and hygienic social norms by population's members.
    
\end{abstract}

\section{Introduction}

The outbreak of COVID-19 pandemics in early 2020, caused by the SARS-CoV-2 virus, represents a challenge for public health and scientific research at world level. Indeed, it has produced an outstanding increase on research about the virus itself and, more in general, epidemics in many scientific fields~\cite{mar20}.

Since the COVID-19 pandemic requires large-scale behaviour change, insights from the social and behavioural sciences have been shown to be crucial to help align human behaviour with the recommendations of epidemiologists and public health experts~\cite{bav20}.
A large literature has found social norms -- the unwritten social rules that regulate behavior in everyday contexts~\cite{bic05,sze21} -- to be crucial in solving these challenges~\cite{gel21,che21,vri22} and determining the behaviour of people during the emergency~\cite{vri22,gol20}.

Apart from research in virology seeking for vaccines and efficacious drugs against COVID-19, an important topic is the analysis of the dynamics of the pandemic. The starting point is the well known SIR (``Susceptible-Infected-Recovered") model and its modifications. Here, we also start from a modified SIR model and, based on the recent literature, we will explore the effect of social norms on the spreading of the virus throughout the population.

\ 

As a first step, in the next subsections we define the model we are going to employ in the following.

\ 

\subsection*{Compartmental Models and SIR(S)} 

The most common way to model an epidemic spreading is to divide the exposed population in classes or compartments, according to the state of the individuals. How these compartments are determined and how the agents change their status following the evolution of the epidemics defines the different models~\cite{pas15}. The most famous and long-standing of these is the SIR model~\cite{ker27}, where the compartments are three: Susceptibles (S), Infected (I) and Recovered (R). The first compartment includes the "healthy" individuals, who have not been infected (yet), infected and recovered are the subsequent stages of the process. An S agent can be infected by an I one with rate $\beta$, an I agent recovers with rate $\gamma$ reaching the R status, which is a frozen status (in SIR, recovered individuals persist in their condition). Therefore, at each time, the state of the system is given by the densities of susceptible and infected individuals, $x=x(t)$ and $y=y(t)$, respectively; the recovered density $z=z(t)$ is given by the normalization: $z=1-x-y$ (if the time scale of spreading of the disease is much shorter than average human life, and neglecting phenomena as migration, it is reasonable to assume the amount of population constant in time). Assuming the mean-field approximation, which is equivalent to consider the population set on a complete graph ($i.e.$, every agent is directly connected with everyone else), the dynamics of the epidemics will be given by the following system of differential equations~\cite{pas15,ker27}:

\begin{equation}
\left\{
\begin{array}{l}
\dot x = -\beta xy \\
\ \\
\dot y = \beta xy -\gamma y \\
\ \\
\dot z = \gamma y \ . 
\end{array}
\right.
    \label{sir}
\end{equation}

\ 

\begin{figure}
  \centering
  \includegraphics[width=91mm]{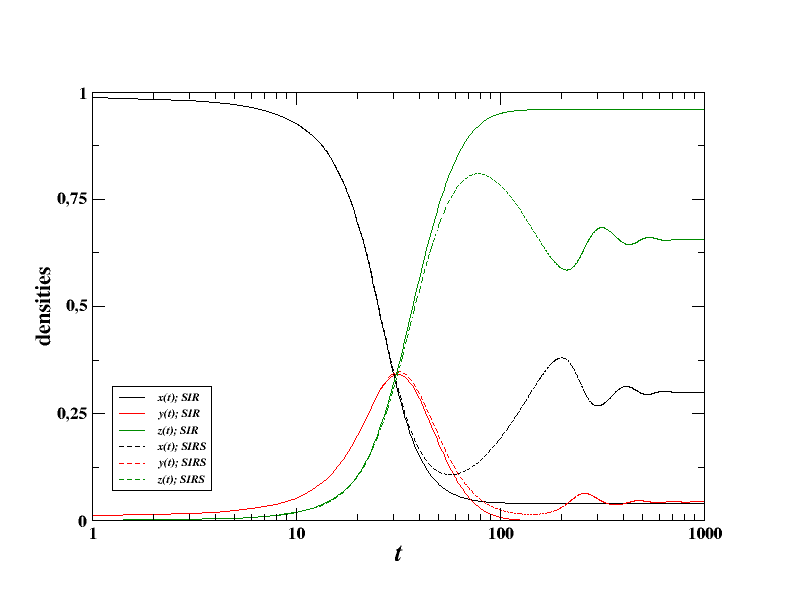}
  \caption{Comparison between time behaviours of SIR and SIRS models: the possibility for recovered individuals to get susceptible again, though at a very slow time scale, changes dramatically the behaviour of the system. The abscissa axis (time) is in logarithmic scale to make the dynamics clearer. Values of the model parameters: $\beta=0.50,\ \gamma=0.15$, and for SIRS $\tau_s=100$. See also Reference~\cite{gon11}.}
  \label{fig1}
\end{figure}

\noindent An explicit solution of the previous system can not be obtained analytically, due to its non-linearity, but it can be easily solved numerically. In short, the main features of the model are the following. First, it is straightforward that in the final state no infected agents are still present, since $y_\infty=0$ is a necessary condition for having equilibrium. Therefore, in the final state there will be a part of the population who has never been infected (let $x_\infty$ be the final density of it), and the remaining part made up by recovered people ($z_\infty=1-x_\infty$). Anyway, in general we observe a non-trivial the dynamics which depends on the precise values of the parameters at stake. Indeed, infected density $y(t)$ may undergo an exponential growth stage before vanishing, or going directly to zero. In particular, the quantity $R=\beta/\gamma$ (called {\it basic reproduction  number}) is crucial to determine the behaviour of $y(x)$: if $R<1$, $y(t)$ vanishes rapidly, if instead $R>1$, there is an initial exponential growth of the number of infected people~\cite{ker27} (in Figure~\ref{fig1} we show an instance of such behaviour).
SIR model turns out to be a useful description of an epidemics when some conditions are fulfilled, as in particular a quick dynamics with respect to average human life and the impossibility for individuals to become susceptible again once recovered.

\ 

In order to broaden the scope of validity of this approach, SIR model has to be suitably generalized. Therefore, let us now consider a modified SIR model, which is better known as SIRS ($i.e.$, ``Susceptible-Infected-Recovered-Susceptible")~\cite{gon11}. As the original one defined by Equations~(\ref{sir}), it models a population whose members can be susceptible to the infection through direct contact among individuals. The compartments are the same, susceptible (S), infected (I), and recovered (R), but in this model recovered individuals become susceptible again after an average time $\tau_s$. Therefore, in this case the system of differential equations describing the dynamics have the form

\begin{equation}
\left\{
\begin{array}{l}
\dot x = -\beta xy + \frac{1}{\tau_s}z\\
\ \\
\dot y = \beta xy -\gamma y \\
\ \\
\dot z = \gamma y - \frac{1}{\tau_s}z \ . 
\end{array}
\right.
    \label{sirs}
\end{equation}

\ 

\noindent Of course, in the limit $\tau_s\rightarrow+\infty$ the recovered individuals are practically immune and the model reduces to the classical SIR. 

SIRS allows a more realistic way to represent the dynamics of infections for which people who has been infected becomes susceptible again after a while, as for example CoVid-19~\cite{che20}. In this model the system can end up to a final, mixed state where all the three kinds of agents are in general non-zero. The possibility of recovered people to get infected again is actually a key point which determines the fate of the system~\cite{gon11}: indeed, while in the SIR case the every possible equilibrium state has no infected ($\lim_{t\rightarrow\infty}y(t)=0$), with finite value of $\tau_s$ we have in general final states with a constant infected rate larger than zero. More importantly, the convergence to the equilibrium presents oscillations of vanishing amplitude, as shown in Figure~\ref{fig1}. This property of the model shows how fundamental is to know whether the infected people get immune once recovered or not. For instance, in reference~\cite{ves20} by Vespignani and coworkers, an evaluation of the effectiveness of the ban of international flights from China is carried out: though the model of the flows among airports is well established, the epidemic model assumes the immunization after recovering, which is the case for SARS-CoV-2 virus and its variants~\cite{che20}. Therefore, models with no possibility of reinfection are useful only on short time scale. Further studies are still needed to clarify definitively the issue and adopt the right class of models.

\subsection{The role of topology}

The results of the SIR and SIRS dynamics summarized above hold in complete graphs (mean-field), that is, all the agents are directly connected with each other. Indeed, realistic epidemics happen in complex topologies, determined by the actual relations among individuals. In general, the strength of the pandemics is at its maximum in complete graphs, because the more the agents are connected, the easier is the infection from an infected to a susceptible agent, and this holds true for most compartmental models~\cite{pas15,goz20}. This feature will be  verified also in the model we are going to define in the following, as we will illustrate in the next sections.

\subsection{Social Norms and epidemic spreading}

The analysis of data from surveys during the different waves of the COVID-19 pandemic shows that people from different countries differ in the way they comply with social norms help limiting the spread of the virus, such as hygiene norms: not only the averages, but also standard deviations and higher order moments vary from country to another~\cite{gel21}. Therefore, we explores how the shape of the distribution of norm compliant behavior influences the dynamics of the contagion. In particular, we focus on the role of the standard deviation and examine situations where everyone adopts the same behaviour compared to others where instead there is a large variability throughout the population (e.g., norm compliance is very high for some individuals and very low for others) may produce very different outcomes, as it has been already shown, for instance, in tax compliant behavior ~\cite{giulia}.

\subsubsection{Heterogeneous infection rates}

The parameters of the classical compartmental models as SIR and SIRS, systems~(\ref{sir}) and~(\ref{sirs}), are constant real numbers: that is, the rates $\beta$ and $\gamma$, and the average recovery time $\tau_s$ are considered uniform throughout the population and constant in time. This is in many cases a useful approximation, but here we want to refine it and consider the individual variability in space and time: since we are interested in the normative influence over the spreading of the pandemics, it is convenient to start considering the infection rate $\beta$ as non-uniform throughout the population.

\ 

\section{The Model}

In all compartmental models, originally the infection rate is uniform: that is, the probability if susceptible to get the infection when interacting with an already infected agent is the same for everyone. In general, this is not the case, for both physiological~\cite{dav20} and behavioural reasons~\cite{gel21,han16}. In the real world every individual has his/her own $\beta$. As a first refinement of the model, we consider the parameter $\beta$ as an individual one: therefore, we define at the beginning the distribution $\{\beta_i\}_{i=1,\dots,N}$. Since in this case every agent has a different infectious rate, when a susceptible individual $a$ meets an infected one $b$, the probability that $a$ is in its turn infected will be the geometric average of the two infection rates: $\beta_{b\rightarrow a}= \sqrt{\beta_a\beta_b}$. This because an individual which respects perfectly all the hygienic measures should not, ideally, infect nor be infected.

In order to check the effect of the heterogeneity of the infectious rate throughout the population, we will accomplish simulations where, fixed the average $\langle\beta_i\rangle$ and the values of the remaining parameters, the only change will be the variance of the distribution $\{\beta_i\}_i$. In this way, we will be able to sort out the role of fluctuations of the infectivity in the dynamics of the epidemic spreading.

\subsection{Symmetric distributions}

Let us start by considering only symmetric distributions of $\beta$, that is, since $\beta\in[0,1]$, distributions where the probability density is such that

\begin{equation}
        F(\beta)=F(1-\beta) \ . 
\end{equation}

\noindent Therefore, we will also have $\langle\beta\rangle=1/2$. We consider the following distributions with average $1/2$:

\begin{itemize} 
    \item Delta distribution centered in $\beta=1/2$, that is, every agent as infectivity ratio equal to 1/2. Therefore, the variance is $\sigma^2=0$.
    \item Uniform distribution in the real interval $[0,1]$ ($\sigma^2=1/12$).
    \item A derived distribution given by a Gaussian centered in 0 and transformed by this rule:
    $$
    \mathcal{P}(x) = \frac{1-\tanh\left(\mathcal{G}_\Sigma(x)\right)}{2}
    $$
    where $\mathcal{G}_\Sigma(x)$ is a Gaussian distribution with variance $\Sigma$.
    \item A bimodal distribution such that the only possible values of $\beta$ are $x_0\in[0,1/2)$ and $1-x_0$, each with probability $1/2$; the variance is here $\sigma^2=1/4-x_0(1-x_0)$.
\end{itemize}

We checked the behaviour of the model keeping fixed all the parameters at stake but the variance of the distribution $\{\beta_i\}_i$. 

\subsubsection{Results}

As already stated above, we performed a set of simulations in order to single out any possible effect on dynamics due to the variance of the distribution of $\beta$. Therefore, we focused first on different distributions with the same mean value, then we considered the very same distribution with different values of $\sigma^2$.

\begin{figure}
  \centering
  \includegraphics[width=91mm]{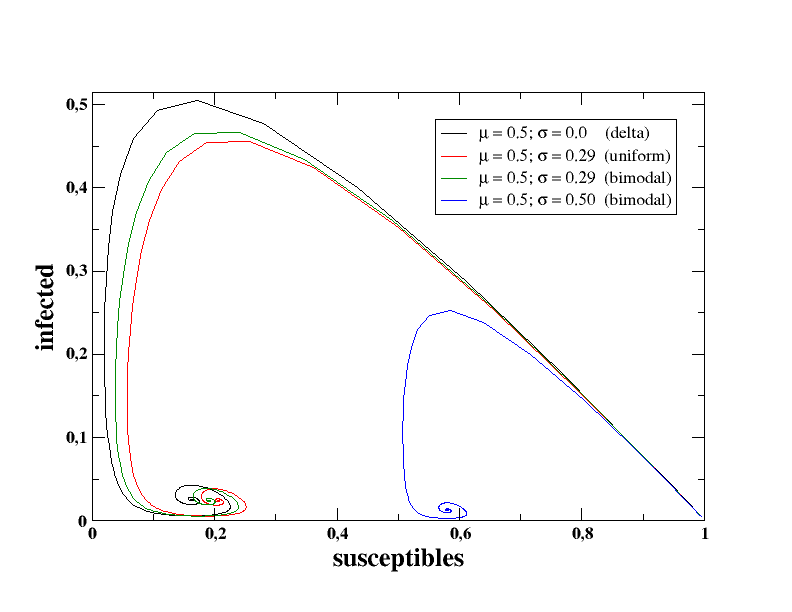}
  \caption{Comparison of the dynamics of the SIRS ABM model among different distributions of $\beta_i$, being fixed the mean value $\langle\beta\rangle=0.5$ and the remaining parameters of the model ($N=2\times10^4$, $\gamma=0.15$, $\tau=200$). Distributions utilized: delta, bimodal, uniform.}
  \label{fig4}
\end{figure}

In Figure~\ref{fig4} we show the time behaviour of the model in the $I$-$S$ plane (i.e., we show the orbits in the space defined by the infected and susceptible densities), for different distributions and $\langle\beta\rangle=0.5$. As it is easy to discern, as the variance increases, the maximum of the infected ratio decreases: that is, having fixed the average, the variability of the distribution helps the system resist the pandemics. The effect of higher-order moments is very small, since the difference between uniform distribution and a bimodal with the same variance is barely perceptible. 

\begin{figure}[ht]
  \centering
  \includegraphics[width=91mm]{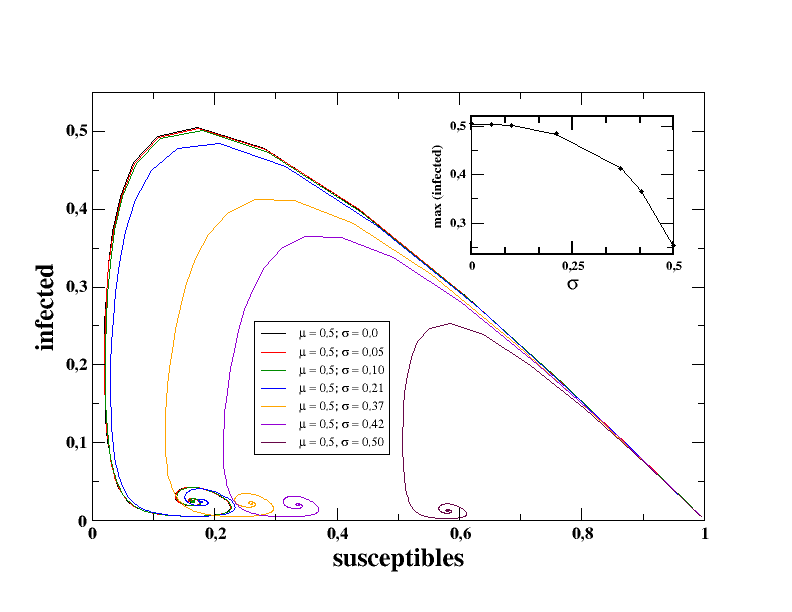}
  \caption{Comparison of the dynamics of the SIRS ABM model among $\tanh$-Gaussian distributions of $\beta_i$ with different variance, being fixed the mean value $\langle\beta\rangle=0.5$ and the remaining parameters of the model ($N=2\times10^4$, $\gamma=0.15$, $\tau=200$). {\bf Inset:} Behaviour of the maximum value of infected ratio as a function of the standard deviation $\sigma$ for $\tanh$-Gaussian distributions.}
  \label{fig5}
\end{figure}

In Figure~\ref{fig5}, on the other hand, we consider only $\tanh$-distributions: again, the effect of changing variance is quite strong, especially in the limit $\sigma\rightarrow0.5$: in the inset of the same figure, we further clarify this point by showing the maximum level of infection during the whole dynamics as a function of $\sigma$. 


\subsection{Theoretical analysis}

As clearly demonstrated in the previous section, the heterogeneity of the distribution $\{\beta_i\}_i$ affects heavily the dynamics of the epidemics and the level of damage it can reach. In particular, it results that, fixed the average, higher heterogeneity implies less global infectivity. We can figure out the mechanism behind this phenomenon if we consider the limiting case of the extreme bimodal distribution with $x_0=0$: in this case, half population has exactly $\beta=0$, that is, all these individuals cannot infect nor being infected at all, they are practically isolated from the rest of individuals. Therefore, even though half population is heavily subject to the infection, the presence of such "invisible to the contagion" agents makes the system much more resistant than a population where everyone has $\beta=0.5$.

\ 

It is possible to justify also analytically the previous conclusions. For simplicity, let us take into consideration a bimodal distribution of $\{\beta_i\}$ with mean value $\langle\beta\rangle=1/2$, defined as follows:

\begin{equation}
    \beta_i =
\left\{
\begin{array}{rlr}
\beta_+=\frac{1}{2}+\varepsilon & \mbox{with probability } \frac{1}{2} & \  \\
\ & \ & \  \\
\ & \ & \ \ \  \  \varepsilon\in\left(0,\frac{1}{2}\right] \\
\ & \ & \  \\
\beta_-=\frac{1}{2}-\varepsilon & \mbox{with probability } \frac{1}{2} & \ 
\end{array}
\right.
\label{binom_d}
\end{equation}

\ 

\noindent whose standard deviation is indeed $\sigma=\varepsilon$. Now, we can split the populations in two subpopulations with higher and lower infection rate, respectively, so that each density is also split: $x_{\pm},\ y_{\pm}, z_{\pm}$, where of course we have $x_++x_-=x$, $y_++y_-=y$, $z_++z_-=z$, being the densities with the subscript $\pm$ referring to the agents with $\beta=\beta_{\pm}$. Therefore, each differential equation of the system~(\ref{sirs}) doubles: for example, the dynamics of infected will be described by the following pair of differential equations:

\begin{equation}
\left\{
\begin{array}{l}
\dot y_+ = \beta_+x_+y_++\sqrt{\beta_+\beta_-}\ x_+y_- -\gamma y_+ \\
\ \\
\dot y_- = \sqrt{\beta_+\beta_-}\ x_-y_++\beta_-x_-y_+ -\gamma y_- \ . 
\end{array}
\right.
    \label{split_sirs}
\end{equation}

\ 

\noindent At the beginning of the dynamics we can assume $y_{\pm}\ll x_{\pm}\simeq\frac{1}{2}$, so that previous system can be rewritten in a more compact form as follows:
$$
    \dot y = \dot y_+ + \dot y_- \simeq -\gamma y + \frac{1}{2}\left(\frac{1}{2}+\varepsilon\right)y_+ + \frac{1}{2}\left(\frac{1}{2}-\varepsilon\right)y_- + \frac{\sqrt{1-\varepsilon^2}}{4}y \ , 
$$

\noindent which yelds

\begin{equation}
\dot y = \gamma\left[R(\varepsilon) - 1\right]y + \frac{\varepsilon}{2}(y_+-y_-) \ , 
\label{split1}
\end{equation}

\ 

\noindent where 

\begin{equation}
R(\varepsilon) \equiv \frac{1+\sqrt{1-\varepsilon^2}}{4\gamma} \ . 
\label{barR}
\end{equation}

\ 

\noindent At the very early stages of the dynamics we can assume the last term at right side of Equation~(\ref{split1}) negligible with respect to the first one: at $t=0$ the initial cluster of infected individuals is made up indifferently by agents of both kinds, because at the beginning of the contagion nobody is aware of the risk and behavioural norms to cope with it are not implemented yet, so that it is reasonable to pose $|y_+-y_-|\ll y$ (notice that the assumption becomes exact in the limit $\varepsilon\rightarrow0^+$). Therefore, reminding that the quantity $\varepsilon$ is simply the standard deviation of the distribution defined in Equation~(\ref{binom_d}), we finally get

\begin{equation}
    \dot y = \gamma\left[R(\sigma) - 1\right]y \ . 
    \label{split2}
\end{equation}

\ 

\noindent Equation~(\ref{split2}) is formally the same as in the homogeneous case, but here the basic reproduction number depends on the standard deviation of the distribution $\{\beta_i\}$. As it is easy to understand from Equation~(\ref{barR}), $R(\sigma)$ is a decreasing function of its argument, meaning that the higher the heterogeneity of the distribution is, the less the epidemics spreads throughout the system.

\subsection{Effects of topology}

\begin{figure}[ht]
\begin{centering}
\includegraphics[width=0.515\textwidth]{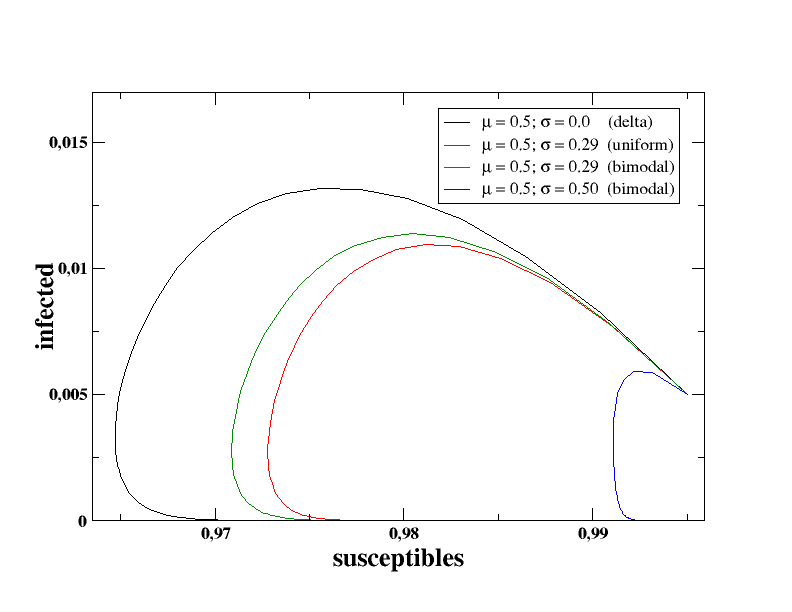}\qquad
\includegraphics[width=0.515\textwidth]{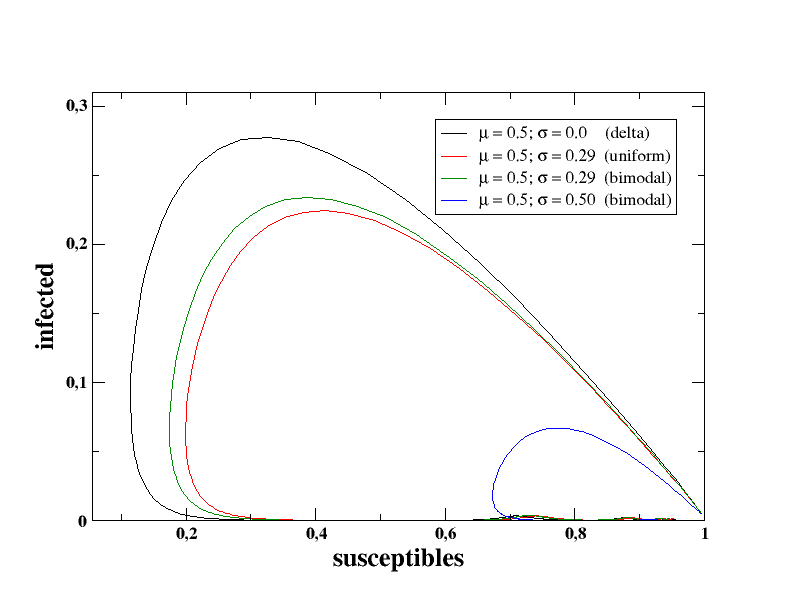}
\end{centering}
\caption{
Comparison of the dynamics of the SIRS ABM model among different distributions of $\beta_i$, being fixed the mean value $\langle\beta\rangle=0.5$ and the remaining parameters of the model ($N=2\times10^4$, $\gamma=0.15$, $\tau=200$). Distributions utilized: delta, bimodal, uniform. {\bf Left:} annealed Small-World network; {\bf Right:} one-dimensional ring.}
\label{steady}
\end{figure}

The role of topology has already heavily studied for SIR-like models~\cite{xia12,gun19,odo21}. Here we aim for a comparison between topology and heterogeneous infectivity effects in SIRS ABM. In general, complex topology can be considered a tool to model top-down measures to control the epidemic spreading (limiting physical contacts among individuals is equivalent to change the network in which the population is embedded); on the other hand, the heterogeneity of the infection rates accounts for the individual differences in physiological characteristics (how prone one is to get and/or transmit the pathogen) and the variability in complying social norms about health and hygiene (the more agents observe the norms, the less is likely they can infect or being infected), as we have already illustrated in previous sections. Indeed, as it is clearly shown in Figure~\ref{steady}, the qualitative behaviour of the model is the same (the larger the variance, the better is the system's response to the epidemics), but the share of the population infected is much smaller.

\ 

\section{Time heterogeneity}

Another strong assumption usually adopted by compartmental models is the time invariability of the parameters involved, in particular the infection rate $\beta$. In this section we want to take into consideration the possibility of non-constant rates. Initially, we neglect the distributions throughout the population, so we are back to the delta-distribution case.

In References~\cite{tak21a,tak21b} some real data about epidemics CoVid-19 waves in Japan are fitted by considering the so-called Avrami~\cite{avr39,avr40} equation for describing the time behaviour of the infected density. The original Avrami equation can be written in the form~\cite{pri65}

\begin{equation}
    y(t) \propto \exp(-Kt^m) \ , 
    \label{avrami}
\end{equation}

\noindent where $m>1$ and $K$ is the fitting parameter. Historically, Avrami equation was firstly written down to describe phase transitions at constant temperature, but the Equation~(\ref{avrami}) can be obtained by the SIR/SIRS equation for the infected density at initial times (when $x\simeq1$) with both $\beta$ and $\gamma$ proportional to the same power of time:

\begin{equation}
    \dot y = \beta_0t^n\cdot y -\gamma_0t^n\cdot y = (\beta_0-\gamma_0)\cdot t^ny \ , 
    \label{sir_avrami}
\end{equation}

\noindent where $n\in\mbox{\bf R}^+$, and whose solution is actually

\begin{equation}
y(t) = y_0\cdot\exp\left(\frac{\beta_0-\gamma_0}{n+1}t^{n+1}\right) \ . 
\label{sir_avr_sol}
\end{equation}

\noindent It is clear that assuming the very same time behaviour of $\beta$ and $\gamma$, and that they start from zero and increase with time, is quite unrealistic. In fact, to our purposes we have to characterize better the time dependence of the rates. A more realistic choice for time-dependent $\beta$ and $\gamma$ is the following:

\begin{equation}
\gamma = \mbox{constant}; \ \ \  \beta(t) =
\left\{
\begin{array}{rlr}
\beta_0-\beta_1t^n & t\le T & \  \\
\ & \ & \  \\
\ & \ & \ \ \  \  \beta_0\in(0,1];\ \ \ 0<\beta_0-\beta_1 T\ge0 \\
\ & \ & \  \\
\beta_0-\beta_1T^n & t > T & \  
\end{array}
\right.
    \label{assump}
\end{equation}

\ 

\noindent where $T\le\sqrt[n]{\beta_0/\beta_1}$ is the time needed to reach the minimum value of the infection rate.
Here we have split $\beta$ in a constant part plus a variable one, which takes into account the modifications of the infection rates due to the change in agents' behaviour. For simplicity we set $\beta_1>0$, that is, we assume that the emergence of the epidemics make people behave more cautiously decreasing the infection rate. The recovery rate is left constant in time, since it is less influenced by the norm-induced behaviour of the agents and it is reasonable to assume that it varies much more slowly with respect to $\beta$. Inserting relations~(\ref{assump}) in the equation for the infected density at initial times, for $t<T$ we obtain

\begin{equation}
    \dot y(t) = (\beta_0-\beta_1t^n)y(t) - \gamma y(t) = \gamma_0(R_0-1)y(t) - \beta_1 t^ny(t) \ , 
    \label{mod_avr_eq}
\end{equation}

\noindent whose general solution is

\begin{equation}
    y(t) = y_0\cdot\exp\left[\gamma(R_0-1)t-\frac{\beta_1}{n+1}t^{n+1}\right] \ , 
    \label{mod_avr_sol}
\end{equation}

\noindent being $R_0\equiv\beta_0/\gamma$. Equations~(\ref{mod_avr_eq}) and~(\ref{mod_avr_sol}) reduce correctly to their classical SIR/SIRS versions in the limit $\beta_1\rightarrow0^+$.

It is straightforward to see that for $R_0<1$ the solution~(\ref{mod_avr_sol}) is always decreasing for $t\ge0$, so that the infection gets rapidly reabsorbed. On the other hand, if $R_0>1$ the time behaviour of the infected density becomes much more interesting. As a matter of fact, $y(t)$ is initially increasing, up to the instant $t^*$ after which it decreases. Such critical time is

\begin{equation}
    t^* = \left(\frac{|R_0-1|}{R_1}\right)^{\frac{1}{n}} \ , 
    \label{crit_t}
\end{equation}

\noindent where we defined $R_1\equiv\beta_1/\gamma$. Of course, it must be $t^*<T$, otherwise $y(t)$ remains increasing. It is easy to understand that $t^*$ actually exists if $\beta(T)=\beta_0-\beta_1T<\gamma$, that is, if $R(t)\equiv\beta(t)/\gamma$ manages to become smaller than the recovery rate. Finally, it must be reminded that these results hold while the approximation $y(t),\ z(t)\ll1$ is still true.

\ 


\section{Perspectives}
\label{concl}

In this work we have tested the effect of paved the way for a possible approach able to link the social norms dynamics to the epidemic spreading dynamical models, so that the socio-behavioural features of a population can be taken properly into account together with the physiological ones. As a first step, we focused on the contagion rate from infected to susceptible individuals by considering its variability throughout the population (and also in time). Indeed, once we set every population member with his/her own infection rate, this personal value will be determined also by the degree to which individuals obey to preventive social norms, ({\it i.e.}, compliance with hygienic norms, quarantines, etc.), which then will be taken into account in the study of the evolution of the contagions.

A reliable mathematical model describing with good approximation the main characteristics and dynamics of a pandemic, like the COVID-19 one, cannot be reached without accurate experimental and observational data about how people in different countries have responded to the emergency, in particular about how much their behaviour has been compliant with the hygienic and health social norms. In the future, suitable, interdisciplinary studies are already planned to address this need.

\


\newpage 
\begin{thebibliography}{50}

\bibitem{mar20}
Already up to April 2, 2020, only in the arXiv 264 preprints, come out since January, had "COVID-19" or "CoVid-19" in their title and/or abstract. A graphics showing the increasing of research in epidemics after the outbreak of the pandemics, up to the end of February, can be found at
\newline https://medium.com/@tomaspueyo/coronavirus-the-hammer-and-the-dance-be9337092b56 (chart 10).


\bibitem{bav20}
Bavel, J. J. V., Baicker, K., Boggio, P. S., Capraro, V., Cichocka, A., Cikara, M., \dots \& Willer, R. Using social and behavioural science to support COVID-19 pandemic response. Nature human behaviour, 4(5), 460-471 (2020).

\bibitem{bic05}
Bicchieri, C. The grammar of society: The nature and dynamics of social norms. Cambridge University Press  (2005).

\bibitem{sze21}
Szekely, A., Lipari, F., Antonioni, A., Paolucci, M., S\'anchez, A., Tummolini, L., \& Andrighetto, G. Evidence from a long-term experiment that collective risks change social norms and promote cooperation. Nature communications, 12(1), 1-7 (2021).

\bibitem{gel21}
Gelfand, M. J., Jackson, J. C., Pan, X., Nau, D., Pieper, D., Denison, E., \dots \& Wang, M. The relationship between cultural tightness–looseness and COVID-19 cases and deaths: a global analysis. The Lancet Planetary Health, 5(3), e135-e144 (2021).

\bibitem{che21}
Chevallier, C., Hacquin, A. S., \& Mercier, H. COVID-19 vaccine hesitancy: Shortening the last mile. Trends in cognitive sciences, 25(5), 331-333 (2021).

\bibitem{vri22}
Vriens, E., Tummolini, L., Andrighetto, G. (under review). Social norm interventions to decrease vaccine hesitancy (2021).

\bibitem{gol20}
Goldberg, M., Gustafson, A., Maibach, E., van der Linden, S., Ballew, M. T., Bergquist, P., \dots \& Leiserowitz, A. (2020). Social norms motivate COVID-19 preventive behaviors.

\bibitem{pas15}
Pastor-Satorras, R., Castellano, C., Van Mieghem, P., \& Vespignani, A. Epidemic processes in complex networks. Reviews of modern physics, 87(3), 925 (2015).

\bibitem{ker27}
Kermack, William Ogilvy, and Anderson G. McKendrick. "A contribution to the mathematical theory of epidemics." Proceedings of the royal society of London. Series A, Containing papers of a mathematical and physical character 115.772 (1927): 700-721.

\bibitem{gon11}
Gon\c{c}alves, Sebasti\'an, Guillermo Abramson, and Marcelo FC Gomes. "Oscillations in SIRS model with distributed delays." The European Physical Journal B 81.3 (2011): 363.

\bibitem{che20}
Chen, Wen-Hsiang, et al. "The SARS-CoV-2 Vaccine Pipeline: an Overview." Current Tropical Medicine Reports (2020): 1-4. See also: https://youtu.be/SzckrIwDeI0 (in italian).

\bibitem{ves20}
Chinazzi, Matteo, et al. "The effect of travel restrictions on the spread of the 2019 novel coronavirus (COVID-19) outbreak." Science (2020).

\bibitem{goz20}
Gozzi, N., Scudeler, M., Paolotti, D., Baronchelli, A., \& Perra, N. (2020). Self-initiated behavioural change and disease resurgence on activity-driven networks. arXiv preprint arXiv:2011.03754.

\bibitem{giulia}
Andrighetto, G., Zhang, N., Ottone, S., Ponzano, F., D'Attoma, J.,  \& Steinmo, S. (2016). Are some countries more honest than others? Evidence from a tax compliance experiment in Sweden and Italy. Frontiers in psychology, 7, 472.

\bibitem{dav20}
Davies, N. G., Klepac, P., Liu, Y., Prem, K., Jit, M., \& Eggo, R. M. (2020). Age-dependent effects in the transmission and control of COVID-19 epidemics. Nature medicine, 26(8), 1205-1211.

\bibitem{han16}
Han, D., Li, D., Chen, C., \& Sun, M. (2016). How the heterogeneous infection rate effect on the epidemic spreading in activity-driven network. International Journal of Modern Physics C, 27(06), 1650057.

\bibitem{xia12}
Xia, C., Wang, L., Sun, S., \& Wang, J. (2012). An SIR model with infection delay and propagation vector in complex networks. Nonlinear Dynamics, 69(3), 927-934.

\bibitem{gun19}
G\"und\"u\c{c}, S. (2019). A study on the effects of diffusion of information on epidemic spread. International Journal of Modeling, Simulation, and Scientific Computing, 10(03), 1950015.

\bibitem{odo21}
\'Odor, G. (2021). Non-universal power-law dynamics of SIR models on hierarchical modular networks. arXiv preprint arXiv:2103.07419.

\bibitem{tak21a}
Takase, Y. (2021). Analysis of COVID-19 infection waves in Japan by Avrami equation. arXiv preprint arXiv:2109.12472.

\bibitem{tak21b}
Takase, Y. (2021). Analysis of COVID-19 infection waves in Tokyo by Avrami equation. arXiv preprint arXiv:2110.13449.

\bibitem{avr39}
Avrami, M. (1939). Kinetics of phase change. I General theory. The Journal of chemical physics, 7(12), 1103-1112.

\bibitem{avr40}
Avrami, M. (1940). Kinetics of phase change. II transformation‐time relations for random distribution of nuclei. The Journal of chemical physics, 8(2), 212-224.

\bibitem{pri65}
Price, F. P. (1965). Some Comments on the``Avrami''Equation. Journal of Applied Physics, 36(10), 3014-3016.

\end{thebibliography}
\end{document}